\documentclass[pre,aps,twocolumn,showpacs,floatfix,10pt]{revtex4-1}
\usepackage[dvips]{graphicx}
\usepackage{amsmath}
\usepackage{amssymb}
\usepackage[nice]{nicefrac}
\usepackage{color}

\usepackage{epstopdf}
\AppendGraphicsExtensions{.tif}

\begin{document}

\title{Brownian yet non-Gaussian diffusion in heterogeneous media: \\
from superstatistics to homogenization}

\author{E. B. Postnikov}
\email{postnicov@gmail.com}
\affiliation{Department of Theoretical Physics, Kursk State University, Radishcheva st., 33, 305000 Kursk, Russia}
\affiliation{Saratov State National Research University, Astrakhanskaya 83, Saratov 410012, Russia}

\author{A. Chechkin}
\affiliation{Institute of Physics and Astronomy, University of Potsdam, Karl-Liebknecht-Strasse 24/25, 14476 Potsdam-Golm, Germany}

\affiliation{Akhiezer Institute for Theoretical Physics, 
Akademicheskaya Str. 1, 61108 Krakow, Ukraine}

\author{I.M. Sokolov}
\email{igor.sokolov@physik.hu-berlin.de}
\affiliation{Institut f\"ur Physik and IRIS Adlershof, Humboldt Universit\"at zu Berlin, Newtonstra\ss e 15, 12489 Berlin, Germany}

\begin{abstract}
We discuss the situations under which Brownian yet non-Gaussian (BnG) diffusion can be observed in the model of a particle's motion in a random landscape 
of diffusion coefficients slowly varying in space. Our conclusion is that such behavior is extremely unlikely in the situations when the particles, introduced into the system at random at $t=0$, 
are observed from the preparation of the system on. However, it indeed may arise in the case when the diffusion (as described in Ito interpretation) is observed 
under equilibrated conditions. This paradigmatic situation can be translated
into the model of the diffusion coefficient fluctuating in time along a trajectory, i.e. into a kind of the ``diffusing diffusivity'' model. 
\end{abstract}

\maketitle

\section{Introduction}

Many experiments and numerical simulations point onto a wide-spread (if not universal) behavior of displacements of tracers in different complex systems: 
The mean squared displacement (MSD) of the tracers grows linearly in time, 
\begin{equation}
 \langle \mathbf{x}^2 \rangle = 2 d D_0 t
 \label{eq:Fick}
\end{equation}
(with $D_0$ being the diffusion coefficient and $d$ being the dimension of space), like in the normal, Fickian diffusion; the probability density function (PDF) of the 
displacements is, however, strongly non-Gaussian. At first such non-Gaussian
behavior accompanied by the MSD growth which is linear in time was observed in a broad class of materials close
to glass and jamming transitions, such as binary Lennard-Jones mixture \cite{Kob1997}, dense colloidal
hard sphere suspensions \cite{Kegel2000,Weeks2000}, silica melt \cite{Berthier2007}, and bidisperse sheared granular materials \cite{Marty2005}. 

Later it was found, that the PDF in most of these cases  follows 
the exponential (Laplace) pattern
\begin{equation}
 P(\mathbf{x},t) \propto \exp\left( - \frac{|\mathbf{x}|}{l(t)} \right) 
 \label{eq:LaplDis}
\end{equation}
with the parameter $l(t)$ characterizing the width of the distribution \cite{Stariolo2006}.
In complex fluids this dependence was explained by the \textit{dynamic heterogeneity of ensemble of tracers} which results  in the
intermittent nature of particles' trajectories  \cite{Chaudhury2007}, see Ref. \cite{BerthierRMP2011} for the review.   

Later on, a very similar behavior was observed in a large amount of systems of a very different nature. Thus, in Wang et al. \cite{Wang1,Wang2},
such a behavior was observed for the motion of colloidal beads on phospholipid tubes and in entangled actin suspensions, and later in the motion 
of tracer particles in mucin gels \cite{Wagner}, the cases very different from systems discussed above. 
In \cite{Wang1} this type of behavior was termed Brownian yet non-Gaussian (BnG) diffusion. More experimental examples 
are mentioned in \cite{Seno}. for recent simulation results see \cite{Miotto}.

In \cite{Wang2} the behavior was attributed to the \textit{heterogeneity of the medium}. Each tracer moves in the environment with 
its own diffusivity $D$, i.e., the PDF of each tracer's displacement follows
\begin{equation}
 G(x,t|D) = \frac{1}{\sqrt{4 \pi D t}} \exp\left(- \frac{x^2}{4 D t} \right)
 \label{eq:Gauss}
\end{equation}
 (in one dimension, or in the projection on the $x$-direction).
The PDF of displacements in an ensemble of tracers is than given by
\begin{equation}
 P(x,t) = \int_0^\infty G(x,t|D) p(D) dD,
 \label{eq:compound}
\end{equation}
where $p(D)$ is the probabitity density of the distribution of the diffusivities $D$. 
If $p(D)$ is exponential, than $P(x,t)$ is a Laplace distribution. Different diffusivities 
for different members of the ensemble may be connected with the fact that the properties of the medium slowly change in space or in time. 

The heterogeneity can be atrributed to the tracers themselves. \textit{Heterogeneous ensembles of tracers} were invoked in several works.
Thus, analyzing the consequences of the fundamental observation in population biology that individuals of the same species are not
identical, it was shown in \cite{Petrovsky} that e.g., the Maxwell type of the distribution of speeds of flying insects may 
lead to the exponential distribution of diffusivities. Analyzing the movement data
of the parasitic nematodes, Hapca et al. \cite{Hapca} showed that there is a significant variation of an effective diffusion coefficient
within the population, and that the distribution of the diffusion coefficients follows a gamma distribution. 
This again leads to the exponential tails of the displacement distribution. 

In mathematical statistics the procedure given by Eq. (\ref{eq:compound}) is called compounding \cite{Dubey}.
A slightly different procedure formulated in the language of random variables, is associated with the concept of  
generalized grey Brownian motion (ggBm) \cite{Mura1,Mura2,Gianni,Vittoria,Sliusarenko}. For Brownian diffusion characterized by 
the Gaussian displacement distribution, Eq.(\ref{eq:Gauss}), this procedure leads to a similar scheme as simple compounding.  
In physics, such procedures fall under the notion of superstatistics \cite{Beck1,Beck2,BeckCohen}. 

All above means that the BnG diffusion appears in many different physical systems and mathematical models, and, in physics, this is probably not
a monocausal phenomenon. 

Similar effects are also seen in anomalous diffusion \cite{Spakowitz,Metzler,Stylianidou,MetzAcX,Krapf}
which case however is not discussed in the present work.

In many cases the form of the PDF changes at longer times for the normal, Gaussian one, see e.g. \cite{Wang1,Wang2}.
Such change cannot be described within the tracers' heterogeneity model unless the properties of tracers change in time
(which is not the case in the experiments using well-controlled tracers). 
In the models of heterogeneous media, the exponential-to-Gaussian transition in the PDF may stem from slow temporal fluctuations of the diffusion coefficient 
(a \textit{diffusing diffusivity} model \cite{ChuS,Sebastian,Seno,Cherayil,Lano}).
As mentioned in \cite{Hapca} and \cite{Seno}, such a transition may also stem from the \textit{quenched spacial heterogeneity} of the medium.
At short times the motion of a tracer is confined to a domain with some diffusivity $D$, while at longer times 
it travels between the domains with different diffusivities, so that the values of the diffusion coefficients fluctuate
along the trajectory. 

In the present work we consider the motion of tracers in a medium with quenched distribution of diffusion 
coefficients slowly varying in space. 
Thus, at short times different tracers see different diffusion coefficients 
while at long times (and correspondingly large scales) the homogenization sets in, and the motion
of each tracer can be described as taking place in a homogeneous medium characterized by some effective diffusion coefficient $D^*$.
The BnG diffusion implies that the mean diffusion coefficient $D_0$ sampled at short times 
is equal to the effective one $D^*$ in the homogenized regime. The main topic of the present discussion is to find out when this is likely to be the case,
and, if it is case, how does the transition between the short-time Laplace and the long-time Gaussian PDF shapes takes place.

\section{Position-dependent diffusion coefficient}

Let us assume that the BnG diffusion phenomenon in some specific medium is fully due to the spacial heterogeneity of the local diffusion coefficient $D(\mathbf{x})$ 
which is position-dependent. In dimensions higher than $d=1$ the system will be assumed to be isotropic on the average,
which considerably simplifies the further discussion.  In order to observe heterogeneity at times which are not very short (much longer than experimental 
sampling time for the trajectory), the local diffusion coefficient $D(\mathbf{x})$ has to vary only slowly in space. 
The correlation length $\lambda$ of  $D(\mathbf{x})$ is therefore mesoscopic, and the characteristic time of the transition from the short time inhomogeneous (superstatistical) to
the homogenized behavior  is $t_H \sim \lambda^2 / D_0$. 

Therefore, we assume that the particle's motion in our system is described by the Langevin equation 
\begin{equation}
 \dot{\mathbf{x}} = \sqrt{2 D(\mathbf{x})} \mbox{\boldmath $\xi$ \unboldmath}(t)
 \label{eq:Lang}
\end{equation}
with the Gaussian noise $\mbox{\boldmath $\xi$ \unboldmath}(t)$ 
fulfilling $\langle \mbox{\boldmath $\xi$ \unboldmath} \rangle=0$ and $\langle \xi_\beta (t) \xi_\gamma(t') \rangle = \delta_{\beta \gamma } \delta(t-t')$ with
$\beta, \gamma$ denoting Cartesian components. 

The knowledge of the diffusion coefficient $D(\mathbf{x})$ as a function of the coordinates is not enough to uniquely define the properties of this 
diffusion. Systems with exactly the same $D(\mathbf{x})$, the ones described by the Langevin equation Eq.(\ref{eq:Lang}), are described by different 
Fokker-Planck equations (FPEs), depending on what interpretation of the stochastic integral is assumed. 
The corresponding FPEs for a given realization of $D(\mathbf{x})$ read 
\begin{equation}
 \frac{\partial P(\mathbf{x},t)}{\partial t}  = \nabla \left[(1 - \alpha) \nabla D(\mathbf{x}) + D(\mathbf{x}) \nabla \right] P(\mathbf{x},t),
 \label{eq:FPE}
\end{equation}
with the interpretation parameter $\alpha$ taking the values in the interval $0 \leq \alpha \leq 1$. 
The typical interpretations, the Ito, Stratonovich and the H\"anggi-Klimontovich (HK) ones, correspond to $\alpha = 0, 1/2$ and 1, respectively. 
Each of these schemes may appear in different physical situations (see e.g. Ref.\cite{Sokolov}).

The equilibrium concentration profile $n(\mathbf{x}) \propto P(\mathbf{x}, t \to \infty)$ 
corresponding to the vanishing flux in a closed system is given by the solution of the equation
$ (1 - \alpha)  n(\mathbf{x}) \nabla D(\mathbf{x})+ D(\mathbf{x}) \nabla n(\mathbf{x}) = 0$
and reads
\begin{equation}
 n(\mathbf{x}) = C \cdot D^{\alpha -1}(\mathbf{x})
 \label{purediff}
\end{equation}
with the constant $C > 0$ depending on the size of the system and on the number of particles therein.  The HK case is the only one when this profile is flat. 
The Fokker-Planck equation for HK interpretation corresponds to the phenomenological second Fick's law
\begin{equation}
 \frac{\partial n(\mathbf{x},t)}{\partial t} = \nabla \left[ D(\mathbf{x}) \nabla n(\mathbf{x},t) \right].
 \label{eq:HK}
\end{equation}
The Ito interpretation leads to the Fokker-Planck equation  
\begin{equation}
 \frac{\partial n(\mathbf{x},t)}{\partial t} = \Delta \left[ D(\mathbf{x}) n(\mathbf{x},t) \right]
 \label{eq:Ito}
\end{equation}
which, in some cases, better reproduces the experimental results for heterogeneous diffusion and is connected with the continuous time random walk (CTRW) scheme \cite{Milligan}.

In what follows we will estimate the long-time diffusion coefficient for different values of interpretation parameter $\alpha$ under different additional conditions and discuss the 
question, which model is the best candidate for reproducing the BnG behavior. We will moreover present simulation results for systems corresponding to HK and Ito models
with respect to the time-dependence of the diffusion coefficient and the corresponding PDF. 

\section{Sampled diffusion coefficients and local diffusivities}

In heterogeneous media, the situations under equilibrium, and the non-equilibrium ones may show very different properties \cite{ACe,Meroz}. 
The first, equilibrium, situation corresponds to the case when the system (medium, already containing tracers) was prepared long before starting the observation. 
The position of the tracer is monitored from the beginning of observation ($t=0$) on. If there are several tracers in the system,
the ones to observe are chosen at random. In this case the probability density to find a tracer at position $\mathbf{x}$ at $t=0$ is
proportional to the equilibrium density $n(\mathbf{x})$ of tracers at the corresponding position. If $n(\mathbf{x})$ is not
constant, the space is not sampled homogeneously. 
The second, opposite, situation corresponds to the case when the tracers were introduced at random exactly at the beginning of the observation, and the  
diffusivity values at initial particles' positions are sampled according to the distribution of $D(\mathbf{x})$. We will call the first and the second situations the \textit{equilibrium sampling}, 
and the \textit{homogeneous sampling}, respectively. 
In \cite{Wagner}, a moving time averaging over the long data acquisition 
time is used to get both the displacement's PDFs and the MSDs, which assumes that the system has enough time to equilibrate.  In \cite{Wang1} the ensemble average was used, 
but time between preparation and the beginning of the observation was not specified. The situation will be equilibrated if this time is much larger than $t_H$.  

\subsection{Distribution of sampled diffusion coefficients}

At short times, i.e. in the superstatistical regime, the PDF of particles' displacements $p(\mathbf{x},t)$ follows
by averaging the Gaussian PDF of particles' displacements in a patch with local value of the diffusion coefficient 
$D(\mathbf{x})$ over the distribution of these local diffusivities close to the particles' initial positions. 
Therefore, the PDF in the superstatistical regime (when each particle can be considered as moving with its own diffusion 
coefficient) is given by
\begin{equation}
 p(\mathbf{x},t) = \int_0^\infty \frac{1}{(4 \pi d D t)^{d/2}} \exp\left( - \frac{|\mathbf{x}|^2}{2 d D t} \right) p_S(D) dD ,
 \label{Trafo}
\end{equation}
where $p_S(D)$ is the PDF of sampled diffusion coefficients. This PDF is \textit{defined} by Eq.(\ref{Trafo}) and may or may not 
be equal to the one-point PDF $p(D)$ of the random field $D(\mathbf{x})$, depending on the type of sampling (equilibrium or homogeneous) implied in experiment.

Let us take as a ``stylized fact'' that the PDF of displacements at short times has the exponential form
\begin{equation}
 p(\mathbf{x},t) = A(t) \exp(-| \mathbf{x}| / l (t) ),
\end{equation}
with $l(t)$ defining the width of the distribution and $A(t)$ being the normalization constant. Requiring that
\[
 \int p(\mathbf{x},t) d\mathbf{x} = 1
\]
and that
\[
 \int \mathbf{x}^2 p(\mathbf{x},t) d\mathbf{x} = 2 d D_0 t
\]
in $d$ dimensions, we obtain the explicit form of the displacements' PDFs
\begin{center}
\begin{tabular}{ll}
 $\displaystyle p(x,t) = \frac{1}{2 \sqrt{D_0 t}} \exp\left(-\frac{x}{\sqrt{D_0 t}} \right) $& in $d=1$\\
 $\displaystyle p(\mathbf{x},t) = \frac{3}{4 \pi D_0 t} \exp \left(- \sqrt{\frac{3}{2}} \frac{|\mathbf{x}|}{\sqrt{D_0 t}} \right) $ & in $d=2$ \\
 $\displaystyle p(\mathbf{x},t) = \frac{1}{\pi (2D_0 t)^{3/2}} \exp \left(- \sqrt{2} \frac{|\mathbf{x}|}{\sqrt{D_0 t}} \right) $ & in $d=3$.
\end{tabular}
\end{center}
Passing to the Fourier transforms in the spacial coordinate in Eq.(\ref{Trafo}) and taking into account the central (rotational) symmetry of the PDF we get
\[
 \tilde{p}(\mathbf{k},t) = \int_0^\infty \exp\left( - D t k^2 \right) p_S(D) dD,
\]
where $k = |\mathbf{k}|$, and therefore see that $\tilde{p}(\mathbf{k},t)$ is the Laplace transform of $p_S(D)$ taken at the value of the Laplace variable $s= tk^2$, and that $p_S(D)$ thus
follows as the corresponding inverse Laplace transform. The values of $\tilde{p}(\mathbf{k},t)$ are 
\[
 \tilde{p}(\mathbf{k},t) = \left\{
 \begin{array}{ll}
  \displaystyle \frac{1}{1 + D_0 t k^2} & \mbox{in } d=1 \\
 \displaystyle \frac{1}{(1 + \frac{2}{3} D_0 t k^2)^{3/2} } & \mbox{in } d=2 \\
  \displaystyle \frac{1}{(1 + \frac{1}{2} D_0 t k^2)^2} & \mbox{in }  d=3 ,
 \end{array}
 \right.
\]
and therefore $p_S(D)$, following as the inverse Laplace transform of these equations in $s=tk^2$, are: 
\[
  p_S(D) = \left\{
 \begin{array}{ll}
  \displaystyle \frac{1}{D_0} \exp \left(- \frac{D}{D_0} \right) & \mbox{in } d=1 \\ 
  \displaystyle \frac{3^{3/2}}{\sqrt{2 \pi} D_0} \sqrt{\frac{D}{D_0}} \exp \left(- \frac{3}{2} \frac{D}{D_0} \right) & \mbox{in } d=2 \\
  \displaystyle \frac{4}{D_0} \frac{D}{D_0} \exp \left(- 2 \frac{D}{D_0} \right) & \mbox{in } d=3.
 \end{array}
\right.
\]
These distributions possess the form of a Gamma-distribution
\begin{equation}
 p_S(D) =\frac{\beta^\beta}{\Gamma(\beta)} \frac{1}{\overline{D}} \left(\frac{D}{\overline{D}} \right)^{\beta-1} \exp \left(- \beta \frac{D}{\overline{D}} \right)
 \label{eq:Sampled}
\end{equation}
with the shape parameter $\beta=(d+1)/2$ (with $d$ being the dimension of space) and with the mean $\overline{D} = D_0$. 

The first inverse moment of the distribution $\langle D^{-1} \rangle = D_0^{-1} \beta \Gamma(\beta-1)/\Gamma(\beta) = D_0^{-1} \beta/(\beta-1)$, which will be
of use in what follows, diverges in $d=1$ and is finite in higher dimensions:
\begin{equation}
 \langle D^{-1} \rangle = \left\{
 \begin{array}{ll}
  3/D_0  & \mbox{in } d=2 \\
  2/D_0  & \mbox{in } d=3.
 \end{array}
 \right.
 \label{eq:FIM}
\end{equation}

\subsection{Distribution of local diffusion coefficients}

In the superstatistical regime corresponding to short times the particles move in different areas with different local diffusion coefficients
$D(\mathbf{x})$. Since the patches do not have to be sampled homogeneously, the distribution of sampled diffusion
coefficients might differ from such of $D(\mathbf{x})$. 
In situations corresponding to homogeneous sampling, the two PDFs, $p(D)$ and $p_S(D)$ coincide, so that $p(D)$ is given by Eq.(\ref{eq:Sampled}).
The same is true also for HK interpretation under equilibrium sampling, when the concentration profile given by Eq.(\ref{purediff}) is flat.
The situation for other interpretations under equilibrium sampling is different. 

The probability density to find a domain with diffusivity $D$ by picking up
a  particle at random is proportional to the particles' density in the corresponding region and to probability to find a region with 
density $n$ among all regions with given diffusivity, i.e. to $n p(n,D)$, where $p(n,D)$ 
is the joint probability density of $n(\mathbf{x})$ and $D(\mathbf{x})$. Thus
\begin{eqnarray}
p_S(D) &=& \mathcal{N} \int_0^\infty n p(n,D) dn \nonumber \\
&=& \mathcal{N} \int_0^\infty n p(n|D) p(D) dn,
\label{eq:SampGeo}
\end{eqnarray}
where  $\mathcal{N}$ is the normalization constant, and, in the second line, $p(n|D)$ is a probability density of the particle concentration conditioned on the diffusion coefficient in the corresponding
domain.
The normalization constant $\mathcal{N}$ is trerefore given by 
\begin{eqnarray*}
 \mathcal{N}^{-1}  &=& \int_0^\infty \left[ \int_0^\infty n p(n|D) dn \right] p(D) dD \\
 &=& \int_0^\infty \langle n | D \rangle p(D) dD
\end{eqnarray*}
where $\langle n | D \rangle$ in the last line is the first conditional moment of the particle concentration.  This gives us a physical interpretation of the 
normalization constant $\mathcal{N}$: it is the inverse of the concentration averaged over all possible landscapes, 
which coincides with a volume mean $\langle n(\mathbf{x})\rangle$ in the thermodynamical limit. 
Therefore, it is useful to introduce the normalized equilibrium concentration at a position $\mathbf{x}$ 
\begin{equation}
\nu(\mathbf{x}) = \frac{n(\mathbf{x})}{\langle n(\mathbf{x}) \rangle},
\label{Nu1}
\end{equation}
and put down Eq.(\ref{eq:SampGeo}) as
\begin{equation}
 p_S(D) = \int_0^\infty \nu p(\nu|D) p(D) d\nu
 \label{eq:psnu}
\end{equation}
(note that $\mathcal{N} n = \nu$ and $p(n | D) dn = p(\nu | D) d \nu$). 

According to Eq.(\ref{purediff}) the connection between $n(\mathbf{x})$ and $D(\mathbf{x})$ is deterministic, $n(\mathbf{x}) = n(D(\mathbf{x})) = C \cdot D^{\alpha - 1}(\mathbf{x})$.
Therefore the normalized concentration is
\begin{equation}
\nu(\mathbf{x}) = \frac{D^{\alpha-1}(\mathbf{x})}{\langle D^{\alpha-1}(\mathbf{x}) \rangle},
\label{Nu}
\end{equation}
and the conditional probability density $p(\nu|D)$ is given by 
\begin{equation}
 p(\nu | D) = \delta(\nu - D^{\alpha - 1} \langle D^{\alpha - 1} \rangle^{-1}). 
 \label{eq:pnuD}
\end{equation}
Subsituting Eq.(\ref{eq:pnuD}) into Eq.(\ref{eq:psnu}) we get
\begin{equation}
 p_S(D) = \frac{D^{\alpha - 1}}{\langle D^{\alpha-1}\rangle} p(D),
 \label{eq:psDpD}
\end{equation}
with the additional requirement that
\begin{equation}
 \langle D^{\alpha-1}\rangle = \int_0^\infty D^{\alpha-1} p(D) dD.
 \label{eq:Norm2}
\end{equation}
Equations Eq.(\ref{eq:psDpD}) and (\ref{eq:Norm2}) are easily solved by noting that the first one gives the form 
of $p(D)$ up to the normalization constant $\langle D^{\alpha - 1} \rangle$:
\begin{eqnarray}
p(D) &=& \langle D^{\alpha - 1} \rangle  D^{1-\alpha} p_S(D) \nonumber \\ 
&=& \langle D^{\alpha - 1} \rangle \frac{\beta^\beta}{\Gamma(\beta)} D_0^{-\beta} D^{\beta-\alpha} \exp \left(- \beta \frac{D}{D_0} \right)
\label{PDD}
\end{eqnarray}
where in the second line we simply substitute the expression for $p_S(D)$ as following from Eq.(\ref{eq:Sampled}). Requiring the normalization of the l.h.s. we get
\begin{equation}
\langle D^{\alpha - 1} \rangle= \frac{\Gamma(\beta)}{\beta^{\alpha-1}\Gamma(\beta - \alpha +1)} D_0^{\alpha-1}.
\label{NormEq}
\end{equation}
Substituting this expression into Eq.(\ref{PDD}) we obtain
\[
 p(D) = \frac{D_0^{\alpha - \beta -1} \beta^{\beta - \alpha + 1}}{\Gamma(\beta - \alpha + 1)} D^{\beta - \alpha} e^{-\beta \frac{D}{D_0}}
\]
and recognize on the r.h.s. a $\Gamma$-distribution 
\begin{equation}
p(D) =\frac{\beta'^{\beta'}}{\Gamma(\beta')} \frac{1}{\overline{D}} \left(\frac{D}{\overline{D}} \right)^{\beta'-1} \exp \left(- \beta' \frac{D}{\overline{D}} \right) 
\label{eq:PDFloc}
\end{equation}
with the shape parameter $\beta'= \beta - \alpha + 1$ and with a mean
\[
 \overline{D} = \frac{\beta-\alpha+1}{\beta} D_0
\]
different from $D_0$. Therefore the PDF of local diffusion coefficients is again given by a Gamma distribution.
For future convenience we repeat the expressions for the parameters of the distribution of local diffusion coefficients as functions of the dimension of space for homogeneous 
sampling ($p(D) = p_S(D)$ with $p_S(D)$ given by Eq.(\ref{eq:Sampled})) and for equilibrium sampling, Eq.(\ref{eq:PDFloc}), in the table below.

\begin{table}[h!] 
\caption{Parameters of the Gamma distribution \label{Tab:Dist} }
 \begin{tabular}{|c|c|c|} \hline
sampling type & shape parameter & mean \\ \hline
homogeneous & $\beta = \frac{d+1}{2}$ & $\overline{D} = D_0$ \\ \hline
equilibrium & $\beta' = \frac{d+3}{2} - \alpha$ & $\overline{D} = \left[ 1 + \frac{2(1-\alpha)}{d+1} \right] D_0$ \\ \hline
 \end{tabular}
\end{table}

The cumulative distribution function corresponding to the PDFs, Eqs. (\ref{eq:Sampled}) and (\ref{eq:PDFloc}), is
\begin{equation}
  F_z(D) = \int_0^D p(D') dD' = \frac{1}{\Gamma(z)} \gamma \left(z, z \frac{D}{\overline{D}} \right) 
\label{CDF}
\end{equation}
with $z=\beta$ or $z=\beta'$ respectively, and  $\gamma(z,x)$ being the lower incomplete $\Gamma$-function.  
This form will be used for generating diffusivity landscapes in simulations, as discussed Section \ref{Sec:Sim}.

In the HK case and in the Ito case for homogeneous sampling 
the parameters are $\beta =(d+1)/2$ and $\overline{D}=D_0$; in the equilibrated Ito case they are $\beta' = (d+3)/2$ and $\overline{D}=\frac{d+3}{d+1} D_0$.  

Calculating $\langle D^{\alpha -1} \rangle = \int_0^\infty D^{\alpha -1} p(D) dD$ for the case of homogeneous sampling we can put down the expression for $\nu(D)$ which will be 
useful when calculating the effective medium properties:
\begin{equation}
 \nu(D) = \frac{\beta^{\alpha-1}\Gamma(\beta)}{\Gamma(\beta + \alpha -1)} \left( \frac{D}{D_0} \right)^{\alpha-1}.
 \label{eq:nuhom}
\end{equation}
For the case of equilibrium sampling  $\langle D^{\alpha -1} \rangle$ is given by Eq.(\ref{NormEq}), and  
\begin{equation}
 \nu(D) = \frac{\beta^{\alpha-1}\Gamma(\beta - \alpha +1)}{\Gamma(\beta)} \left( \frac{D}{D_0} \right)^{\alpha-1}.
 \label{eq:nueq}
\end{equation}

\section{Long-time behavior: Homogenization}
Now let us consider the long-time behavior of the diffusion coefficient corresponding to its homogenization. At long times particles sample large domains of the system 
and feel some effective diffusion coefficient $D^*$. 

The problem of finding effective large-scale characteristics of an inhomogeneous medium is an old one. The simplest approaches correspond to static, time-independent
problems, and the best-investigated situations are pertinent to the homogenization of the electric conductance, and to mathematically similar cases of homogenization 
of dielectric or magnetic successibility, see e.g. \cite{Beran}. In the language of conductivity the situation is as follows: One considers a large piece of medium (inhomogeneous
conductor), say in a form of a slab, with given boundary conditions for the potential (for example, with two opposite sides connected to a battery by highly conducting
electrodes which are thus kept at constant potential difference $\Delta \phi$) and measures the total current through the system. Locally the current density 
follows the Ohm's law $\mathbf{j} (\mathbf{x}) = \sigma(\mathbf{x}) \mathbf{E} (\mathbf{x})$ giving a linear connection between a solenoidal field $\mathbf{j}(\mathbf{x})$ (for which  
$\mathrm{div} \; \mathbf{j} = 0$) and a potential field $\mathbf{E}(\mathbf{x}) = \mathrm{grad} \; \phi(\mathbf{x})$ (for which in the static case $\mathrm{rot} \; \mathbf{E} = 0$ holds). 
Demanded is the connection between the volume mean current density $\overline{\mathbf{j} }= \frac{1}{V} \int_V \mathbf{j} (\mathbf{x}) d \mathbf{x}$ 
and mean electric field $\overline{\mathbf{E}} = \frac{1}{V} \int_V \mathbf{E} (\mathbf{x}) d \mathbf{x}$
in the thermodynamical limit $V \to \infty$: $\overline{\mathbf{j}} = \sigma^* \overline{\mathbf{E}}$, where $\sigma^*$ is the effective conductance.
The mean values of the current density and of the electric field are immediately connected with the total current through the 
system and to the potential difference provided by a battery. A review of methods for calculating $\sigma^*$ for a variety of cases is given
e.g. in Ref. \cite{Sahimi}. 

Standard procedures of finding $\sigma^*$ can be applied to finding the effective long-time diffuson coefficient in the HK case, but not in any case with inhomogeneous equilibrium.
The reason is that the standard phenomenological Fick's equation corresponding to the HK case, Eq. (\ref{eq:HK}), can be considered as a combination of a local continuity
equation $\frac{\partial}{\partial t} p(\mathbf{x}) = - \mathrm{div}\; \mathbf{j(\mathbf{x})}$, with $\mathbf{j}(\mathbf{x})$ being now the probability or particles' flux, 
and the linear relation between $\mathbf{j(\mathbf{x})}$ and an obviously potential field $\mathrm{grad}\; p(\mathbf{x})$,  namely the first Fick's law 
$\mathbf{j} (\mathbf{x}) = D(\mathbf{x}) \; \mathrm{grad} \;p(\mathbf{x})$ serving as an analogue of the Ohm's law. At long times the probability distribution
spreads, and the process of diffusion slows down, so that $\frac{\partial}{\partial t} p(\mathbf{x}) = - \mathrm{div}\; \mathbf{j(\mathbf{x})} \to 0$,
and the methods used in the time-independent electrical case can be safely applied.

In the cases corresponding to inhomogeneous equilibrium the methods cannot be applied immediately. Although the corresponding diffusion equations 
can still be considered as a combination of a local continuity equation $\frac{\partial}{\partial t} p(\mathbf{x}) = - \mathrm{div}\; \mathbf{j(\mathbf{x})}$ with
some linear response law, the last one does not have the form of the Ohm's law. For example, for the Ito case, Eq.(\ref{eq:Ito}), we have
$\mathbf{j} (\mathbf{x}) =  \mathrm{grad} \; D(\mathbf{x}) p(\mathbf{x})$.

The homogenization of the diffusion coefficient in a heterogeneous medium is still similar to the one for the electric conductance \cite{Beran,Sahimi}, 
but with an important difference discussed below. 

Let us assume that we are able to calculate the effective conductance $\sigma^*$ of an inhomogeneous system with local conductances $\sigma(\mathbf{x})$ 
and denote this as a special type of an average, the homogenization mean, by $\sigma^* = \langle \sigma(\mathbf{x}) \rangle_H$. 
Then the effective diffusion coefficient in the homogenized regime for the general diffusive case is given by the same type of the average:
\begin{equation}
D^* = \frac{\langle D(\mathbf{x}) n (\mathbf{x})\rangle_H}{\langle n (\mathbf{x}) \rangle} =  
\langle D(\mathbf{x}) \nu(\mathbf{x}) \rangle_H =\langle \kappa(\mathbf{x}) \rangle_H,
\label{HomAv}
\end{equation}
where $\kappa(\mathbf{x}) = D(\mathbf{x}) \nu (\mathbf{x})$. 
This result follows by considering the stationary flow through a large piece of the medium with concentrations kept constant at
its boundaries \cite{Camboni}. An alternative approach can follow the lines of \cite{Dean} where some specific situations were considered. 
Here we give a simple explanation, not following the lines of the proofs but giving a physical intuition.

The explanation relies on the property $\langle a \sigma(\mathbf{x}) \rangle_H = a \langle \sigma(\mathbf{x}) \rangle_H$ which is evident 
from the initial definition of $\sigma^*$ as connecting the mean current density with the mean electric field.
Let us assume that the diffusing particles carry a charge $q$, and move in electric field $\mathbf{E}$ under the force $\mathbf{f} = q \mathbf{E}$. 
Then the local current density is $\mathbf{j}(\mathbf{x}) = q^2 n(\mathbf{x}) \mu(\mathbf{x}) \mathbf{E}(\mathbf{x})$, with 
$\mu(\mathbf{x}) = D(\mathbf{x})/k_B T$ being the local mobility, giving $\sigma(\mathbf{x}) =  q^2 D(\mathbf{x})n(\mathbf{x}) /k_B T$.
Now we calculate the large scale conductance $\sigma^* = \langle \sigma(\mathbf{x}) \rangle_H$ and assume that it is connected
with the homogenized diffusion coefficient and the mean particles' density by the same Nernst-Einstein relation $\sigma^* = q^2 D^* \langle n(\mathbf{x}) \rangle /k_B T$.
Cancelling all constant prefactors we get $\langle D(\mathbf{x}) n(\mathbf{x}) \rangle_H = D^* \langle n(\mathbf{x}) \rangle$, which is our Eq.(\ref{HomAv}).

The homogenized conductance possesses the upper and the lower bound, from which the Wiener bounds, see e.g. \cite{Beran}, are the most general ones \cite{Bounds}. In our case they read
\begin{equation}
 \left \langle \kappa^{-1}(\mathbf{x}) \right \rangle^{-1} \leq \langle \kappa(\mathbf{x}) \rangle_H \leq \langle \kappa(\mathbf{x}) \rangle.
 \label{Wiener}
\end{equation}
The averages in the bounds in Eq.(\ref{Wiener}) can be considered either as volume means or averages over landscapes. 
In $d=1$ the homogenization mean corresponds \textit{exactly} to the lower bound
\begin{equation}
 \langle \kappa(x) \rangle_H = \left \langle \kappa^{-1}(x) \right \rangle^{-1},
 \label{oned}
\end{equation}
which for conductance follows immediately from the Ohm's law.
 
\begin{figure}
\centering%
\includegraphics[width=\columnwidth]{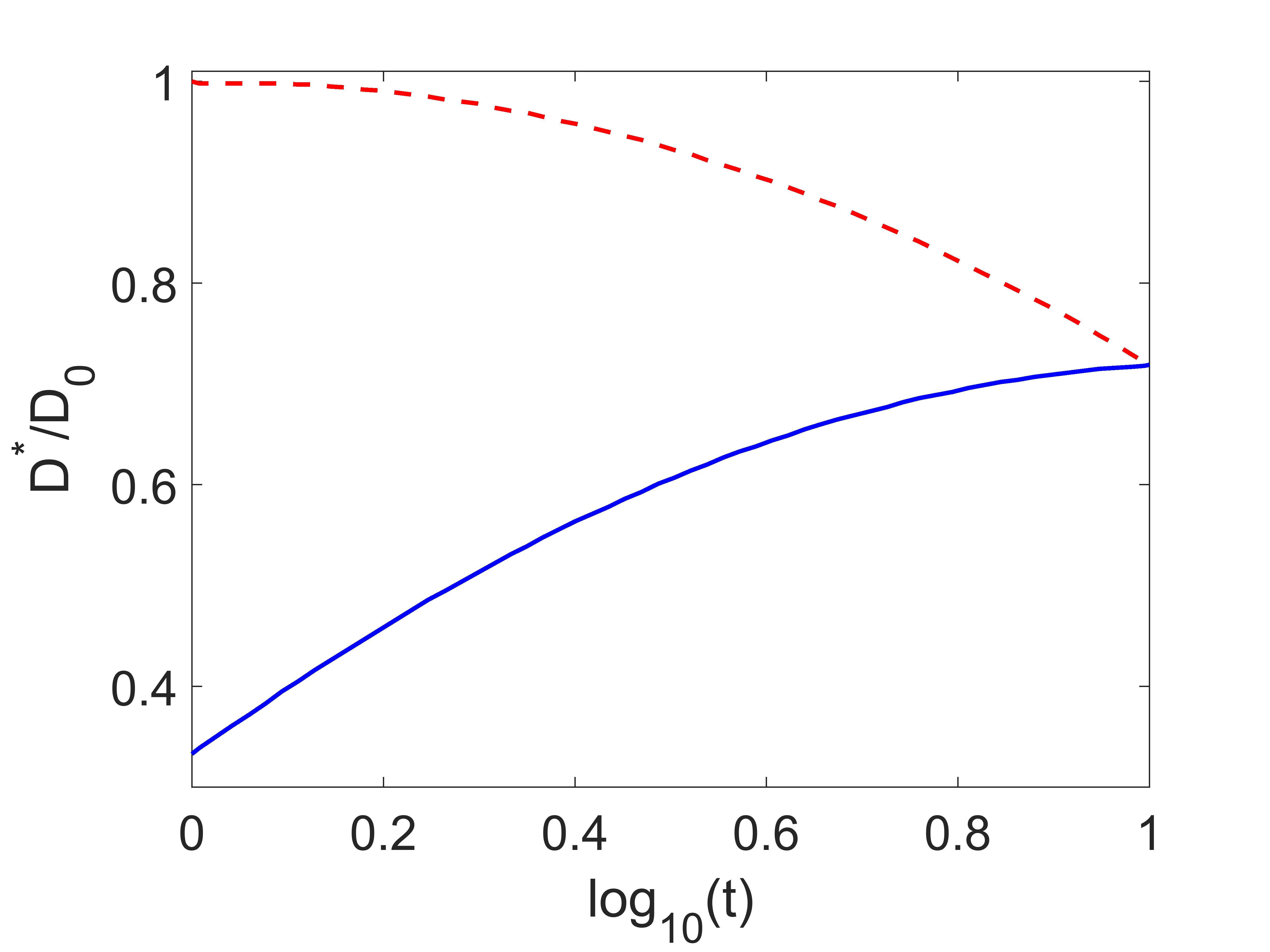}%
\caption{The EMA predictions for the values of $D^*/D_0$ in $d=2$ as a function of the parameter $\alpha$ defining the interpretation. 
The upper (dashed) curve represents the results for the equilibrated case, the lower (full) one for the case of homogeneous sampling.
\label{fig_EMA}}
\end{figure}

Far from percolation transition $\sigma^*$ is typically well reproduced by the effective medium approximation (EMA), 
see \cite{Sahimi} for the discussion. Within this approximation  $D^* = \langle \kappa \rangle_H$ is given by the solution of the equation 
\begin{equation}
 \left \langle \frac{D^* - \kappa}{(d-1)D^* + \kappa} \right\rangle = 0,
 \label{EMA}
\end{equation}
where the average again can be considered either as a volume average or as an average over the distribution of $\kappa$. 
We note that for $d=1$ EMA reproduces the exact result, Eq.(\ref{oned}). 
Eq.(\ref{EMA}) is pertinent to quadratic and cubic lattices in $d=2$ and 3 \cite{Kirkpatrick}, and to continuous $d$-dimensional systems \cite{Sahimi}. 
For known $p(D)$ and $\nu(D)$ Eq.(\ref{EMA}) takes the form 
\begin{equation}
 \int_0^\infty \frac{D^* - D \nu(D)}{(d-1)D^* + D \nu(D)} p(D) dD = 0
\end{equation}
and can be solved numerically by using the parameters of $p(D)$ given in Table \ref{Tab:Dist} and $\nu(D)$ given by Eq.(\ref{eq:nuhom}) and Eq.(\ref{eq:nueq}) for 
homogeneous and equilibrium sampling, respectively. The results of such numerical solution for $d=2$ are shown in Fig. \ref{fig_EMA}.

We note that for the Ito cases, $\alpha = 0$, Eqs.(\ref{eq:nuhom}) and (\ref{eq:nueq}) give $\nu(D) \propto D^{-1}$, so that 
the values of $\kappa = D \nu(D)$ do not fluctuate. According to these equations we have $\kappa = D \nu(D)=\frac{\beta-1}{\beta} D_0$  for homogeneous sampling
(so that $\kappa = \frac{1}{3} D_0$ in $d=2$ and $\kappa = \frac{1}{2} D_0$ in $d= 3$), and $\kappa = D_0$ 
for equilibrium sampling in any dimension. Therefore the corresponding predictions of EMA for Ito cases are essentially \textit{exact}.

\subsection{Systems with homogeneous equilibrium.} For systems with homogeneous equilibrium (the HK interpretation, or the random diffusivity model of Ref. \cite{Dean}) $\nu = 1$ and  there is no 
difference between equilibrium and homogeneous sampling situations: the distributions of $\kappa$ and of $D$ coincide.
The upper Wiener bound corresponds to $\overline{D} =D_0$.
In $d=1$ the first inverse moment $\int_0^\infty D^{-1} p(D) d D$  of the PDF Eq.(\ref{eq:Sampled}) diverges, and the effective diffusion coefficient
$D^*$ given by Eqs. (\ref{HomAv}) and (\ref{oned}) vanishes, giving rise to anomalous diffusion \cite{Camboni}. In $d=2$ and $d=3$ the value of $\langle D^{-1} \rangle$ 
given by Eq. (\ref{eq:FIM}) is finite, and lower Wiener bounds $\langle D^{-1} \rangle^{-1}$ are $D_0/3$ and $D_0/2$, respectively.  
The result of EMA from Eq.(\ref{EMA}) is $D^* = 0.719 D_0$ in $d=2$ and $D^* = 0.852  D_0$ in $d=3$, both smaller than $D_0$. 

Independently on the quality of approximation given by EMA we note that the EMA result is realizable in the continuum case \cite{Milton}: 
the ensemble of all ``disordered'' configurations contains realizations with the effective conductance (diffusivity) equal to the one predicted by EMA.
The lower Wiener bounds are also realizable \cite{Beran}.
Therefore situations leading to the BnG diffusion  are highly unlikely. The ensemble of disordered systems should indeed contain the realizations with diffusivities 
close to the upper bound $D_0$, but also the ones with considerably smaller diffusivities given by the EMA and by the lower bound.
Therefore $D^*$ will typically be lower than $D_0$.

\begin{figure}
\centering%
\includegraphics[width=\columnwidth]{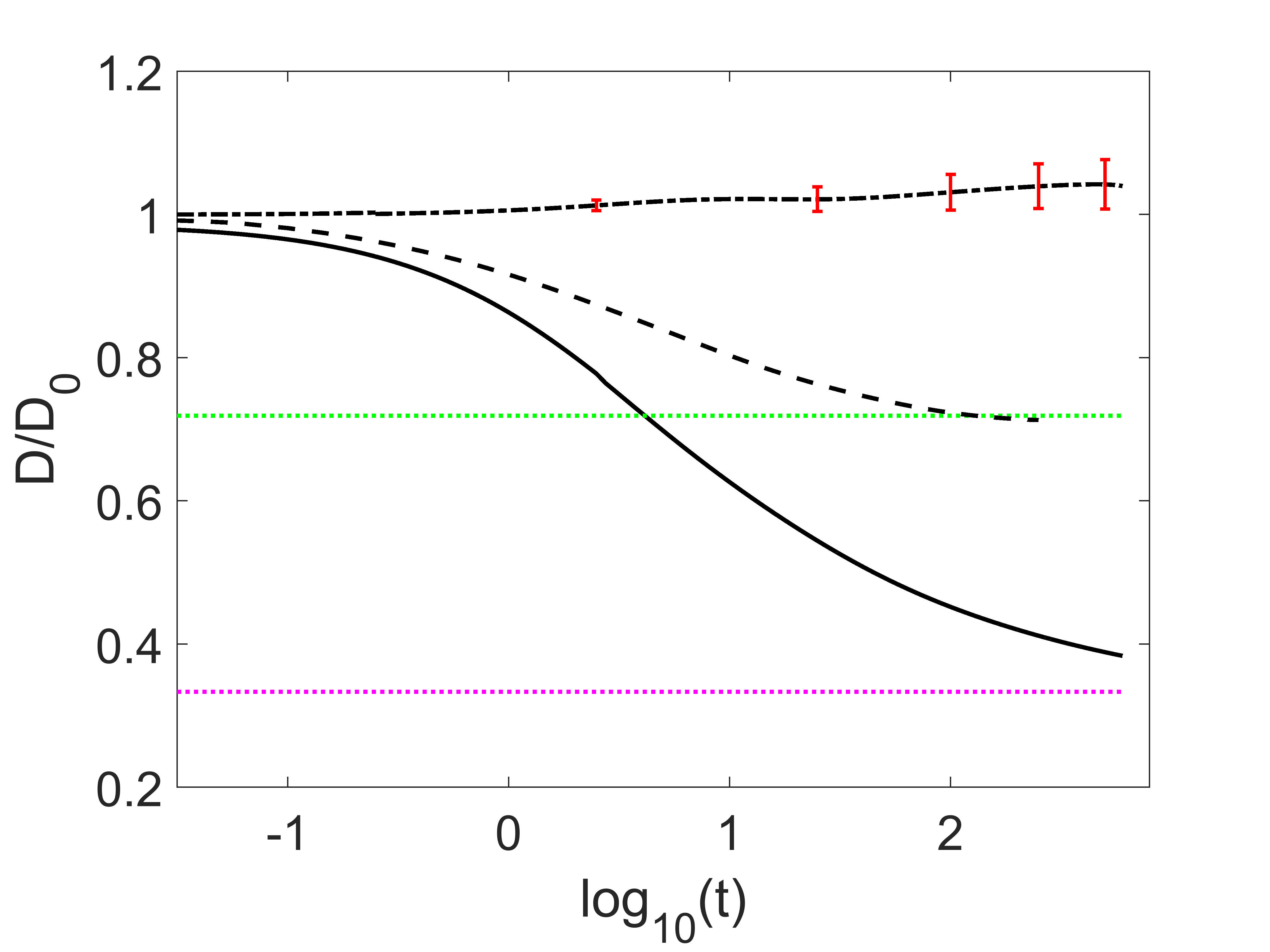}%
\caption{The behavior of $D(t)/D_0$ for the HK interpretation and the two Ito cases for the same values of $\lambda=10$ and $D_0 = 1$ in $d=2$. The upper dashed-dotted line 
corresponds to the equilibrated Ito case and stays horizontal within the statistical error. The error bars show standard deviations of the mean in 1500 realizations.
The lower dashed line gives the time-dependent diffusion coefficient in the HK interpretation. The lowest full line corresponds to the Ito situation under homogeneous sampling. 
For these lines the statistical errors are of the order of the lines' thickness.
The dotted horizontal lines correspond to asymptotic predictions of Fig. \ref{fig_EMA}. 
\label{fig:MSDs}}
\end{figure}

In Fig. \ref{fig:MSDs} we present the full time dependence of the diffusion coefficient in the HK interpretation
as following from numerical simulations. 
The figure shows $D(t) = \frac{1}{4} \frac{d}{dt}\langle |\mathbf{x}|^2(t) \rangle$ normalized on $D_0$ in $d=2$. 
The details of our simulation approach are given in Sec.  \ref{Sec:Sim}.
One readily infers that the diffusion coefficient decays with time, so that no BnG diffusion is observed. 
The value of the terminal diffusion coefficient $D^*$ obtained in simulations agrees well with the EMA prediction.

\subsection{Systems with inhomogeneous equilibrium.}

For systems with inhomogeneous equilibrium the situations under homogeneous and equilibrium sampling are different. We start our discussion using the hints given by the EMA,
as shown in Fig. \ref{fig_EMA}. For homogeneous sampling, we see that the difference between the short time diffusion 
coefficient $D_0$ and the long-time one $D^*$ increases when $\alpha$ decreases from 1. Parallel to the discussion above, there is no reason to await the BnG behavior.
The difference is maximal for the Ito case, when in $d=2$ the terminal diffusion coefficient is exactly one third of the short-time one (in $d=3$ it will make a half of an initial one). 

The result of EMA for equilibrium sampling is shown in Fig. \ref{fig_EMA} as the upper curve. 
We see that the behavior is opposite to the one for homogeneous sampling, and that for $\alpha \to 0$ the difference 
between $D^*$ and $D_0$ vanishes, which result again does not depend on EMA. Moreover,  this behavior persists 
even in $d=1$.  The simulation results for the Ito cases are shown in Fig. \ref{fig:MSDs} along with the one for the HK.

\begin{figure}
\centering%
\includegraphics[width=\columnwidth]{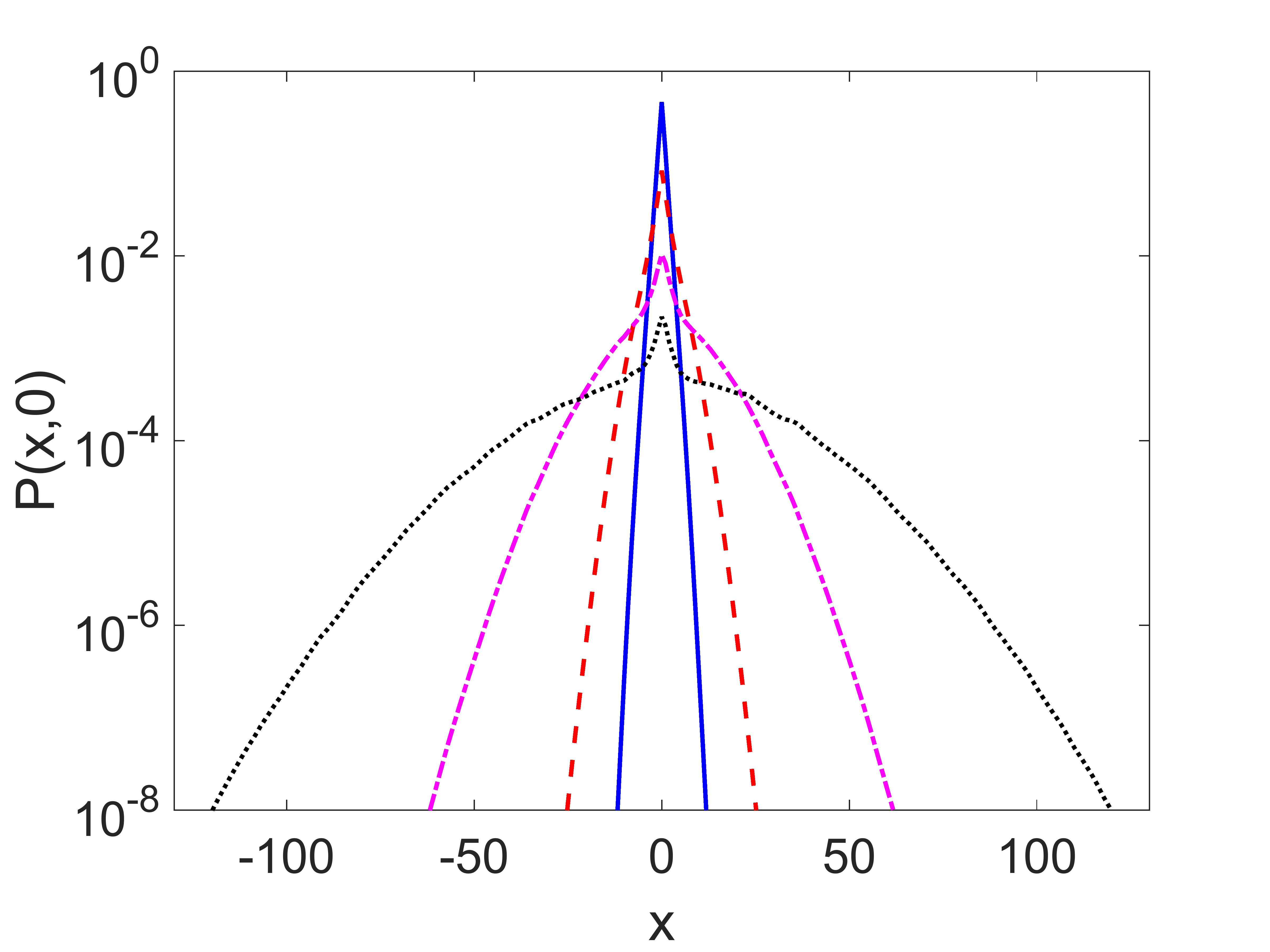}%
\caption{The PDFs $p(x)$ in the equilibrated Ito case in projection on the $x$-axis for $t = 1,10,100$ and 500 in a system with $\lambda=10$ and $D_0 = 1$ in $d=2$, see Sec. \ref{Sec:Sim}
for details of simulations. 
The figure demonstrates the transition from the double-sided exponential to the Gaussian form in a BnG situation. 
\label{fig_Ito_P_equilib} }%
\end{figure}

The effective medium approximation only allows for comparing the initial and terminal values of the diffusion coefficients, but gives 
neither the full time-dependence of the MSD (and therefore of $D(t)$) nor the forms of the corresponding PDFs 
(essentially, no analytical method is known to reliably reproduce such PDFs in the intermediate time domain).
Therefore here we have to rely on the results of numerical simulations. 
Fig. \ref{fig_Ito_P_equilib} displays such PDFs for equilibrated Ito case at different times. These exhibit the transition from exponential to a Gaussian 
distribution, showing a pronounced central peak at intermediate times, which is well known from the experimental realizations \cite{Wang1,Wang2,Wagner}.
 
\section{Simulations of the PDF and MSD in the pure diffusion cases}
\label{Sec:Sim}

In our simulations we start from a discretized model and consider the situation described by the master equation 
\[
  \frac{d}{dt} p_i = \sum_j \left(w_{ij} p_j - w_{ji} p_i \right)
\]
where $i$ and $j$ number the sites of a square or cubic lattice with lattice constant $a=1$. Only the transitions
between neighboring sites are possible. 
This master equation describes a random walk scheme and can be considered as a spatial discretization scheme for the corresponding Fokker-Planck equations, Eq.(\ref{eq:HK}) or Eq.(\ref{eq:Ito}).
The transition rates follow the distribution similar to the distribution of local diffusivities $p(D)$ given by Eq.(\ref{eq:PDFloc}); 
the local diffusivity for the rates which vary slowly in space is simply $D = a^2 w = w$.

For the HK case, Eq.(\ref{eq:HK}), the rates  $w_{ij}=w_{i \leftarrow j}$ satisfy the condition of the detailed balance, i.e., in the absence of the external force $w_{ij}=w_{ji}$ as discussed in Ref. \cite{Sokolov}. 
The rates follow from the PDF $p(w)$ similar to $p(D)$.
To simulate the Ito situations, Eq.(\ref{eq:Ito}), we assume that the transition rates from each site to all its neighbors are the same. 
The transition rate from the site $j$ to any of its neighboring sites $j$ is $w_{ij} = w_j$ and this $w_j$ is distributed according to the corresponding $p(w)$  \cite{Sokolov}. 
The transitions now are asymmetric: $w_{ij} \neq w_{ji}$, which makes a difference. 

To simulate the two situations we generate two or three dimensional arrays of correlated transition rates $w_{ij}$. For the Ito cases all $w_{ij} = w_j$  are defined on the 
sites $j$ of a simple square or simple cubic lattice, corresponding to the lattice of sites at which the probabilities $p_j$ are defined. For the HK case the transition rates
$w_{ij} = w_{ji}$ are defined at the midpoints of bonds of the corresponding lattice of $p_j$. In $d=2$ this lattice of midpoints is a square lattice with the lattice constant equal to $a/\sqrt{2}$ and 
with main axes rotated by $\pi/4$ with respect to the axes of the $p_j$-lattice, but can also be considered as a quadratic lattice with lattice constant $a$ with basis 
(with an additional site placed at a center of a square), which is shifted by $a/2$ with respect to the lattice of $p_j$. 
For three-dimensional case the lattice of the midpoints of bonds of a simple cubic lattice is an octahedral lattice, again considered as a simple cubic lattice 
with basis. Note that the arrays used for simulation of HK and Ito cases are different in size: For example, for simulating a 2d lattice with $N=l \times l$ sites we need $l^2$ 
different values of transition rates for the Ito cases and $2l(l-1)$ (i.e. approximately twice as many) different rates for the HK case.

\begin{figure}[h!]
\centering%
\includegraphics[width=\columnwidth]{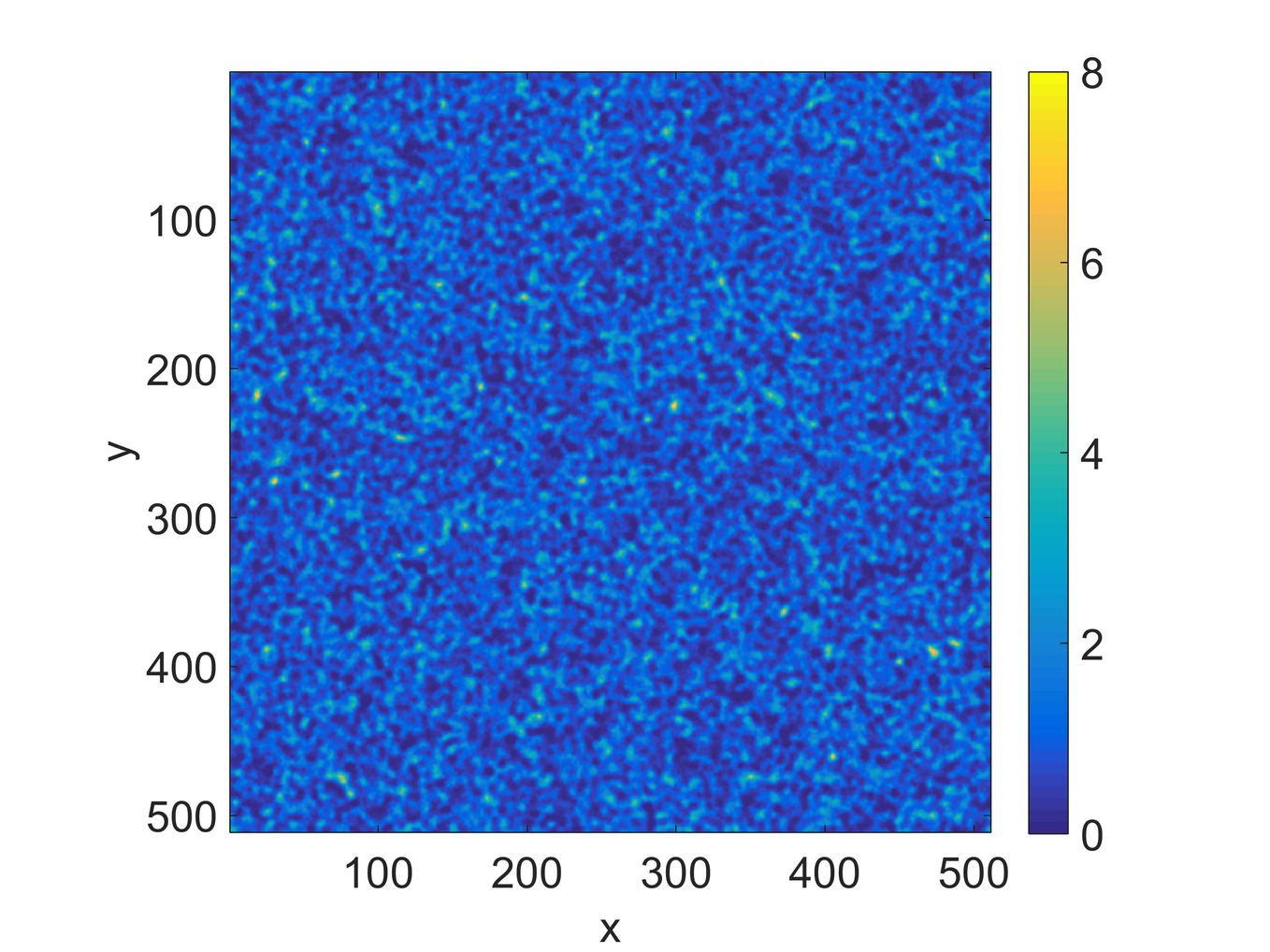}
\caption{A single realization of a landscape of diffusion coefficients (color coded) with $\lambda = 10$ lattice units obtained by the method outlined. \label{fig_landscape} }
\end{figure}

The correlated random variables $w_{ij}$ on the corresponding lattices of transition rates can be easily obtained by a probability transformation. 
Let us call $F^{-1}_{\beta}(y)$ the function inverse to $F_\beta(D)$, as given by Eq.(\ref{CDF}). Then the probability transformation
\[
 w(z) = F^{-1}_{\beta} \left[\frac{1}{2} \mbox{erfc} \left(\frac{z}{\sqrt{2}} \right) \right]
\]
transforms the Gaussian variable $z$ with zero mean and unit variance into a $\Gamma$-distributed $w$ with shape parameter $\beta$ and unit mean. 
The corresponding function $F_\beta(x)$ can be easily inverted (there exists a standard MATLAB implementation for this inverse), 
and therefore the corresponding fields can be easily simulated for any given two-point correlation function. 
In our simulations we use independent Gaussian variables for the uncorrelated case, or a correlated Gaussian landscape with Gaussian correlation function 
$\langle z_i z_j \rangle = \exp(- r^2_{ij}/2\lambda^2)$ with $r_{ij}$ being the distance between the sites $i$ and $j$ on the corresponding lattice, 
with $\lambda$ being the correlation length. Such a correlated Gaussian array is easily obtained by filtering
of the initial array of independent Gaussian variables with their subsequent renormalization necessary to keep $\langle z^2 \rangle = 1$
(note that initial arrays of independent Gaussian variables must be sufficiently larger than its ``internal'' part used in simulations).
The corresponding example of the diffusivity landscape as generated by this method is shown in Fig. \ref{fig_landscape}.

\begin{figure}[h!]
\centering%
\includegraphics[width=\columnwidth]{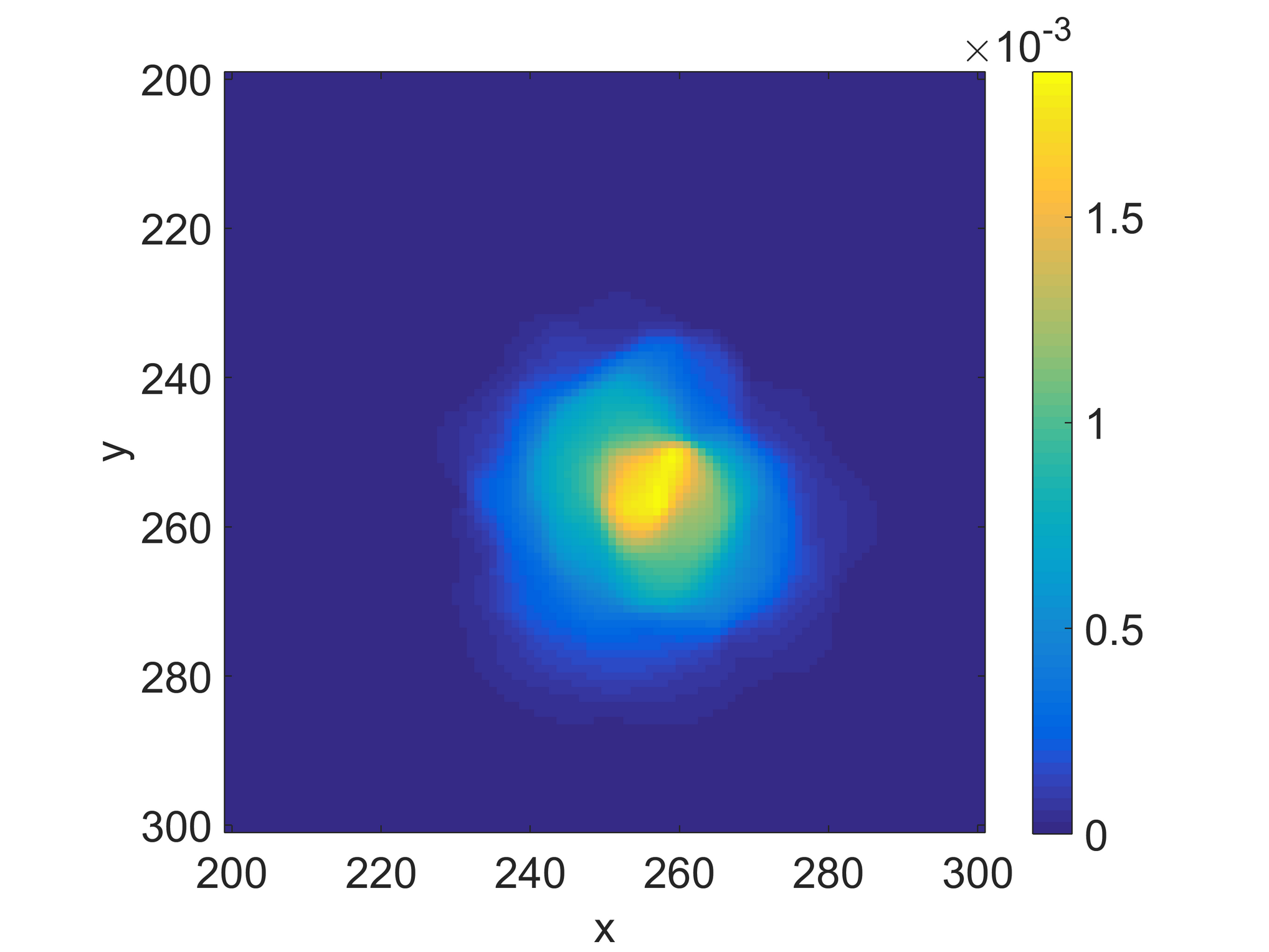}\\
\includegraphics[width=\columnwidth]{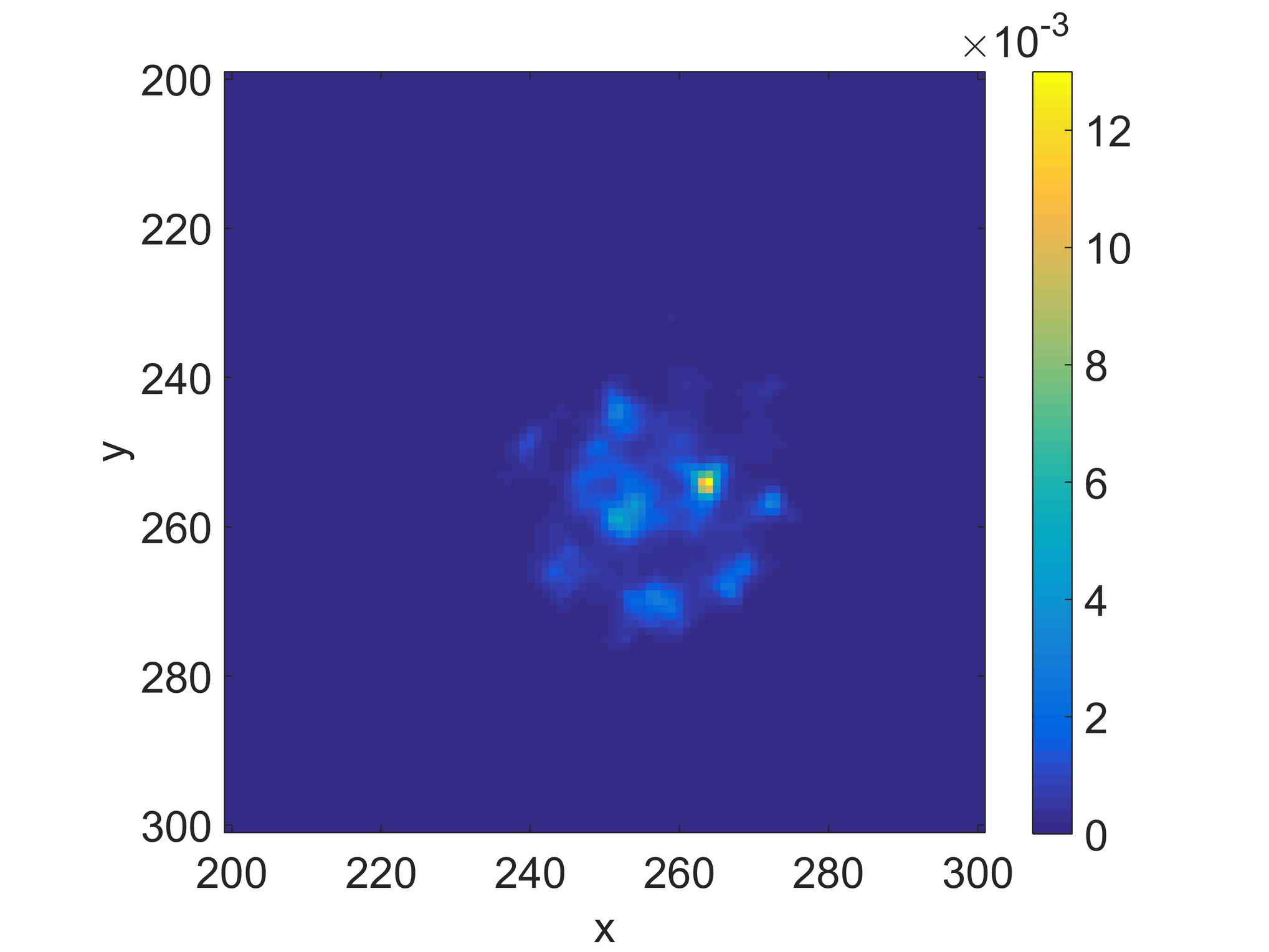}
\caption{Single realizations of probabilities to find particles at corresponding sites of the landscapes with $\lambda=10$ for simulation time $t=0.3$ for the H\"anggi-Klimontovich (upper panel)
and Ito (lower panel) cases. Note the difference in color coding in the two panels. 
One readily infers more or less homogeneous spreading of probabilities for the HK case, and a very granular structure corresponding to probabilities' concentration
in the regions with lower diffusivities for the Ito case. \label{fig_PDFs} }
\end{figure}

\begin{figure}[h!]
\centering%
\scalebox {0.40}{\includegraphics{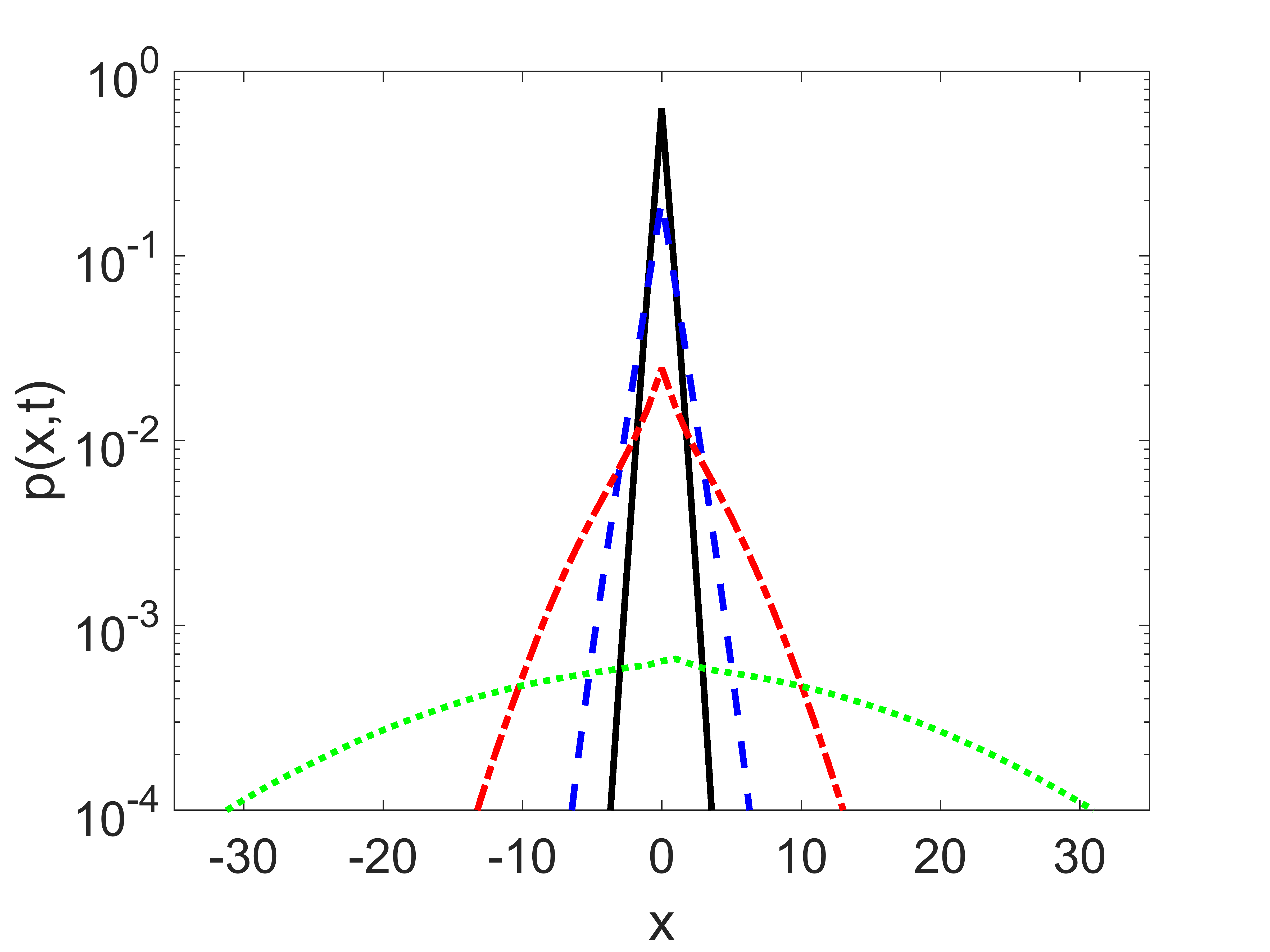}}%
\caption{The probability density functions $p(x,t)$ of the displacements for the HK case with $\lambda = 10$ in $d=2$.
The times are (from top to bottom) $t=0.1, 1, 10$ and 200. Note the logarithmic scale. The figure again clearly demonstrates the transition between the double-sided exponential and the Gaussian form.
Note that the peak at the mode is much less pronounced than in the Ito cases, Fig. \ref{fig_Ito_P_equilib} and Fig. \ref{fig_PDFI}.
\label{fig_PDF}}%
\end{figure}

\begin{figure}[htb]
\centering%
\scalebox {0.40}{\includegraphics{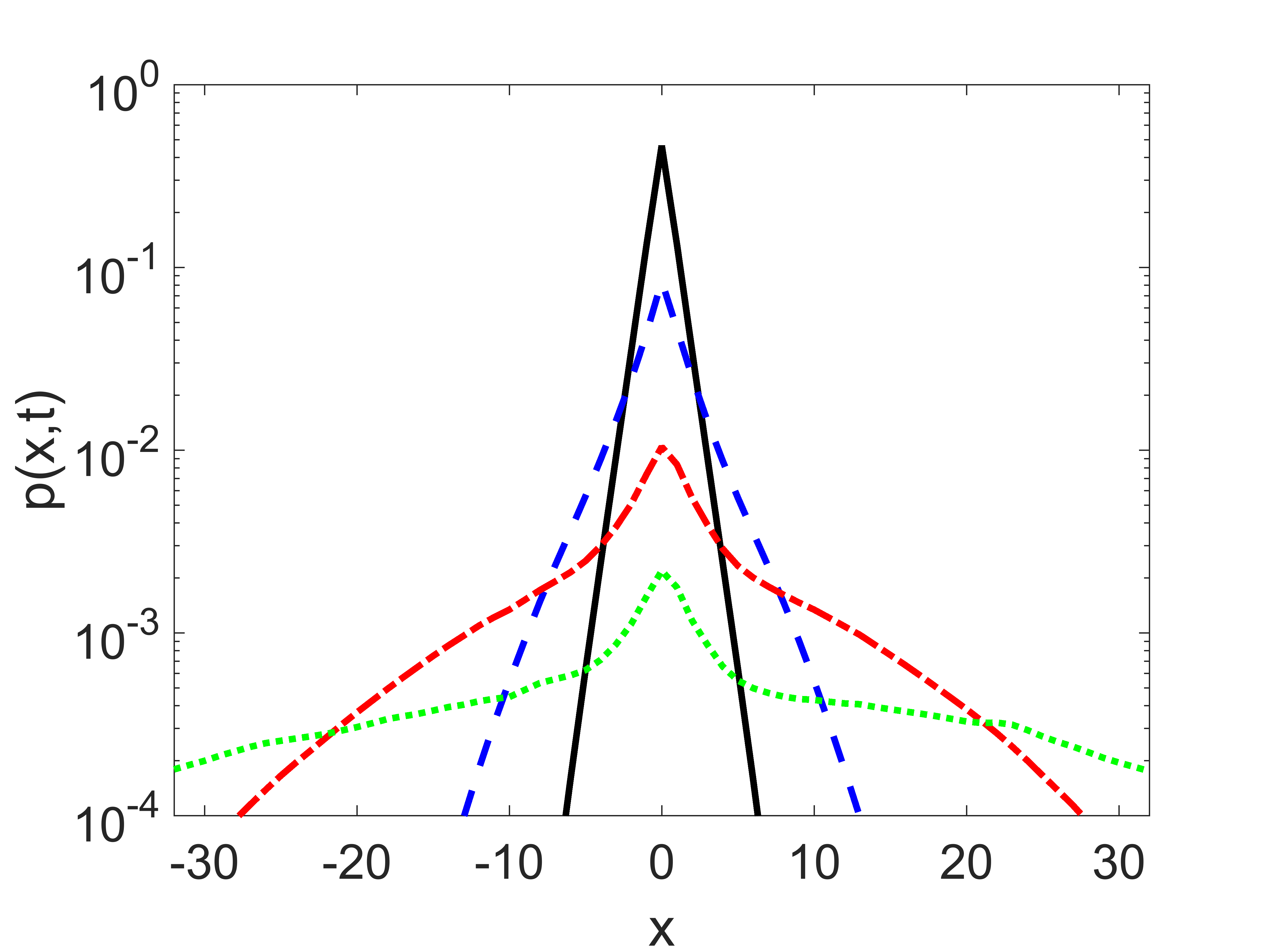}}%
\caption{The probability density functions  $p(x,t)$ for $\lambda = 10$ for the Ito case under homogeneous sampling.
The times are (from top to bottom) $t=1, 10, 100$ and 250. Note the logarithmic scale. 
\label{fig_PDFI}}%
\end{figure}

The simulations of the situations under homogeneous sampling follow by numerical solution of the master equation for a particle starting at the origin.
In 2d the system is of the size $(2L+1) \times (2L+1)$ (i.e. one has $-L \leq k,l \leq L$) where $L=256$ is used in simulations.
The master equation is solved by forward Euler integration scheme to get $p_i(t)$ for each site $i$ characterized by coordinates $\mathbf{r}_i = (k,l)$. 
Fig. \ref{fig_PDFs} shows the exemplary distributions of $p_i(t)$ for the HK and Ito situations, which allow to grasp the differences 
between the cases. Note that the corresponding figures use different realizations of the landscape. 
The probabilities $p_j(t) = p_{(k,l)}(t)$ are then used for plotting the corresponding PDFs. 
The MSD for a particle starting at the origin ($k=l=0$) is given by
$\langle \mathbf{x}^2(t) \rangle = \sum_{k,l = -L}^L (k^2 + l^2) p_{(k,l)}(t)$. 
The PDFs and MSDs are then averaged over different realizations of landscapes $w_{ij}$ (typically  1500 realizations).
In the Ito case under equilibrium sampling the corresponding probabilities and MSDs are weighted with the inverse transition rate from the origin $w_0^{-1}$,
which is proportional to $D^{-1}(\mathbf{0})$ and therefore to the equilibrium concentration at the origin.
Note that since the mean local diffusivity in this case is not equal to $D_0$, additional normalization is applied to keep $D_0=1$.
The PDFs $p(x,t)$ shown in Figs. \ref{fig_Ito_P_equilib}, \ref{fig_PDF} and \ref{fig_PDFI} depict such PDFs for $\mathbf{r}_i = (k,0)$. 

The approach in 3d is exactly the same (except for the fact that $\mathbf{r}_i = (k,l,m)$) but the size of the system is smaller: $L=64$. 

\begin{figure}[h!]
\centering%
\scalebox {0.40}{\includegraphics{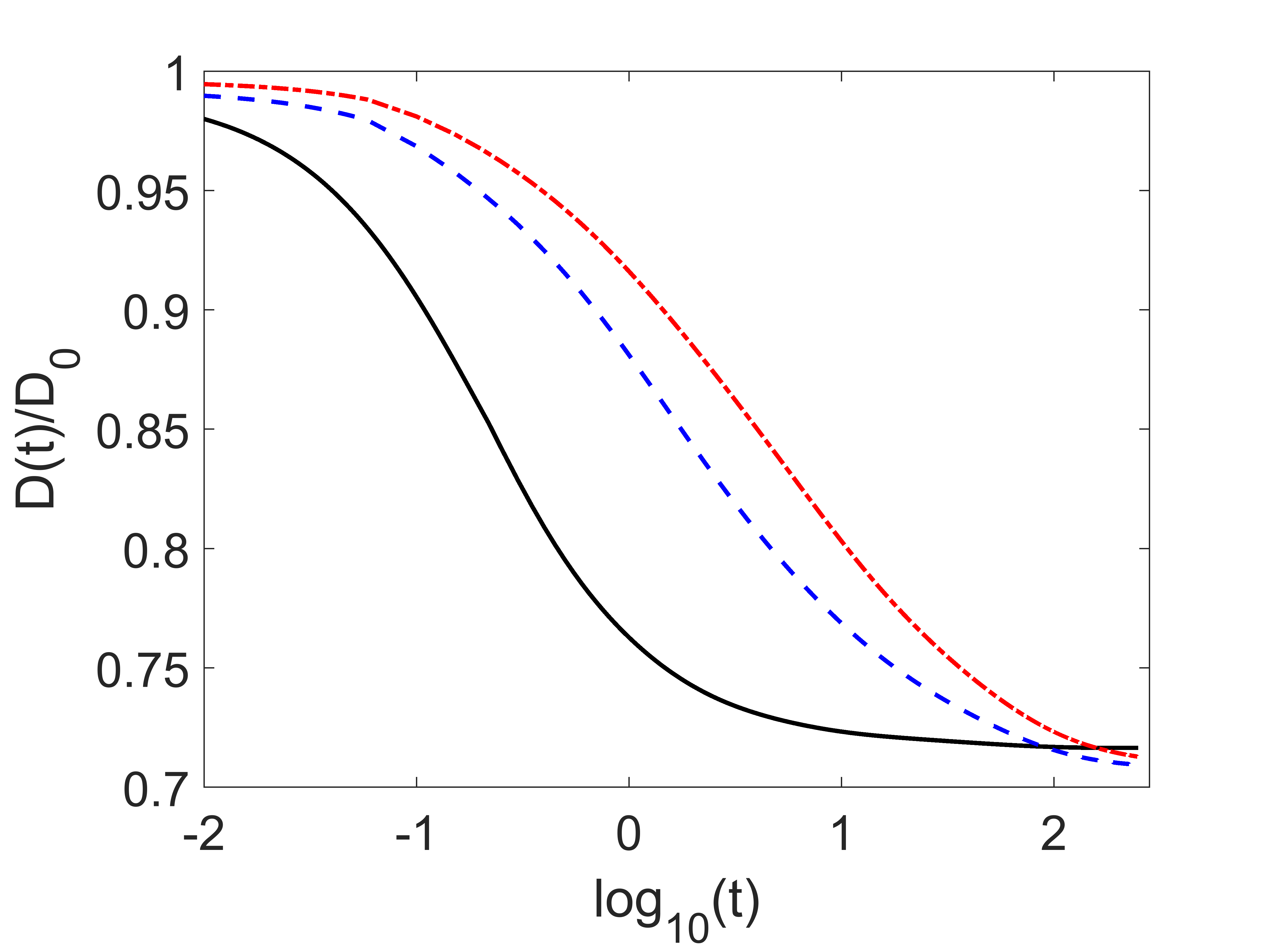}}%
\caption{Time dependent diffusion coefficients in the HK case as obtained by numerical differentiation of MSD for diffusivity landscapes with different correlation lengths in $d=2$:
for uncorrelated transition rates (solid black curve), and for diffusivity landscapes with correlation length $\lambda = 5$ (dashed blue line) and $\lambda = 10$ 
(red dashed-dotted line).
The deviation of the curves in the limit of short times gives the impression about the typical statistical error of simulation. The parameters of simulations imply 
$D_0 =\overline{D} =1$. The simulation results at long times reproduce the ones of EMA within a statistical accuracy (i.e. within a few percent). Thus the terminal diffusion coefficient for the uncorrelated case is $D^*=0.718$
vs. EMA prediction of 0.719, showing the high accuracy of the EMA prediction. The diffusion coefficient is not constant over the time as
it should be in the BnG diffusion. \label{fig_dDdT}}
\end{figure}

\begin{figure}[htb]
\centering%
\scalebox {0.40}{\includegraphics{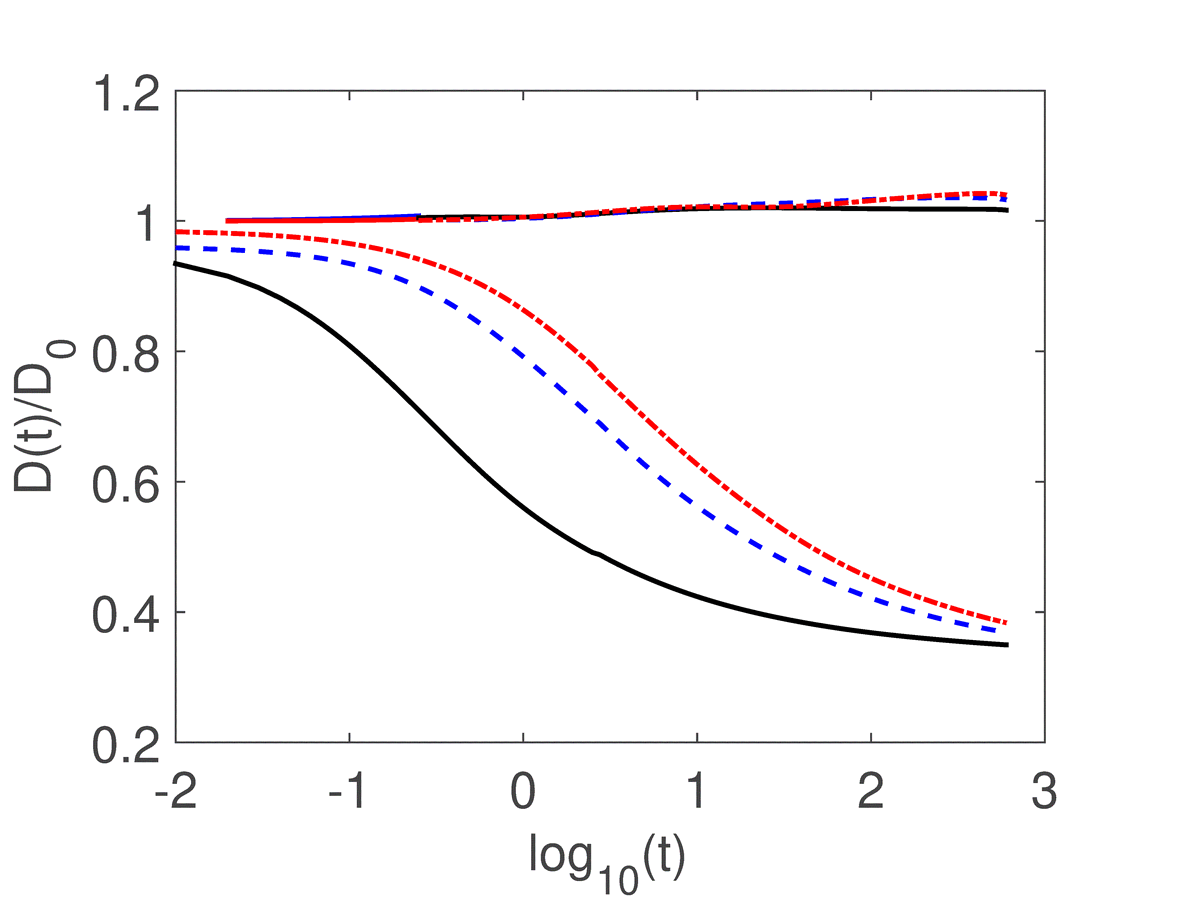}}%
\caption{Time dependent diffusion coefficients in diffusivity landscapes with different correlation lengths in $d=2$
for the Ito case. The lower bundle of curves corresponds to homogeneous sampling and shows the considerable decay of the 
diffusion coefficient with time. The theoretical estimate for the terminal diffusion coefficient is $D^* = \frac{1}{3} D_0 $, and approaching this value by the simulation results is
evident.   The upper bundle of curves corresponds to the equilibrim sampling, with the diffusion coefficient staying constant. The deviations from the constant diffusivity stay within the statistical 
error. The coding of the curves is the same as in Fig. \ref{fig_dDdT}.}
\label{fig_MSDI}%
\end{figure}

\begin{figure}[h!]
\centering%
\scalebox {0.40}{\includegraphics{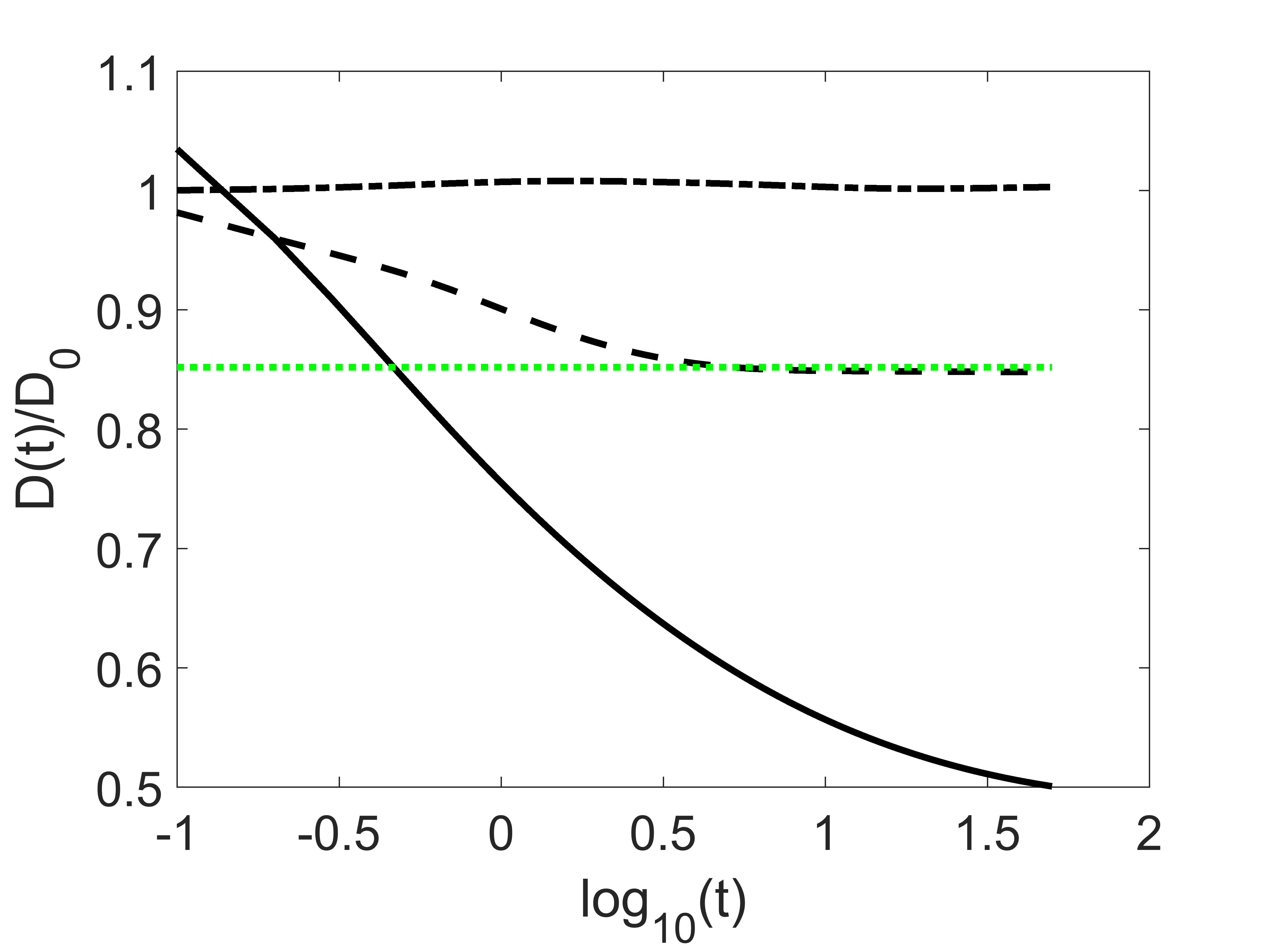}}%
\caption{The behavior of $D(t)/D_0$ for the HK case and the two Ito cases for the same value of $\lambda=3$ in $d=3$. The upper dashed-dotted line 
corresponds to the equilibrated Ito case and stays horizontal within the statistical error. 
The lower dashed line gives the time-dependent diffusion coefficient in the HK case. The lowest full line corresponds to the Ito situation under homogeneous sampling.
\label{fig_3d}}
\end{figure}

The results for time-dependent diffusivity $D(t)$ are obtained by numerical differentiation of 
the corresponding MSD: $D(t) = \frac{1}{2d} \frac{d}{dt} \langle |\mathbf{x} |^2(t)\rangle$. The results for HK case are shown in Fig. \ref{fig_dDdT}. 
Similar results for the Ito case are presented in Fig. \ref{fig_MSDI}. The results shown in Fig. \ref{fig:MSDs} in the previous section are the curve for $\lambda = 10$ from Fig. \ref{fig_dDdT},
and the two corresponding curves for the homogeneous and eqilibrium sampling from Fig. \ref{fig_MSDI}.  To show that the behavior in $d=3$ is similar we present in Fig. \ref{fig_3d}
the results for the time dependent diffusion coefficient in $d=3$ for $\lambda = 3$. The EMA prediction for the HK case corresponds to $D^*/D_0 \approx 0.852$,
and for the Ito case under homogeneous sampling the analytical prediction is $D^*/D_0 = 0.5$.



\section{Discussion}

The properties of diffusion in random diffusivity landscapes strongly depend on the model adopted, but still share some similarities.
The PDFs averaged over the realizations of diffusivity landscapes show similar features in all cases: The gradual transition from exponential to a Gaussian form which happens at the wings
of the distribution while all distributions still show a cusp at the origin at intermediate times. These findings are similar to what is seen in
experiments \cite{Wang1,Wang2,Wagner,Larrat}, and in theoretical models of disordered systems like the trap model of Refs. \cite{Traps,Luo} close in spirit to our Ito model with homogeneous sampling, or a barrier model of Ref. \cite{Stylianidou}, close in spirit to the HK one. The behavior of the MSD in different systems however differs, i.e. may show the BnG diffusion, the crossover between different types of diffusive behavior, 
and even anomalous diffusion.
As we have already seen, within the model class adopted (spatially inhomogeneous systems with slowly varying diffusion coefficient), 
the equilibrated Ito model is the only promising candidate for a model showing the BnG diffusion, i.e. the diffusion coefficient staying constant over the time. 

The Ito interpretation relies on the martingale property, which, in the Gaussian case, means that the increments of the process during small time intervals are symmetric \cite{Stroock}. 
A random walk interpretation of this process can be a continuous time random walk with locally symmetric steps in space, in which the spatial change of the diffusivity is attributed to coordinate-dependent waiting times. 
Such a random walk scheme corresponds to a trap model \cite{Sokolov} which is thus the most prominent candidate for modeling BnG diffusion.
In higher dimensions (in $d=3$ and in $d=2$ approximately, up to logarithmic corrections which are hard to detect) trap models may be mapped to CTRW
under disorder averaging \cite{KlaSo}. CTRW is a process subordinated to a simple random walk (in a continuous limit -- a process subordinated to Brownian motion).
In our case, the waiting times in our CTRW would however be correlated, which is different from the standard CTRW schemes. 
The diffusing diffusivity model \cite{Seno} is also a representative of the class of models subordinated to Brownian motion, and shares some properties with 
the corresponding correlated CTRWs. We note however that this diffusing diffusivity model shows a different 
kind of transition from exponential to Gaussian PDF which does not lead to a cusp at the origin. 

\section{Acknowledgements} EBP is supported by the Russian Science Foundation, project 19-15-00201. AC acknowledges the support by the Deutsche Forschungsgemeinschaft within the project ME1535/7-1.

\end{document}